\documentclass[a4paper]{spie}  

\usepackage{amsmath,amsfonts,amssymb}
\usepackage{graphicx}
\usepackage[colorlinks=true, allcolors=blue]{hyperref}
\usepackage{float}
\usepackage{siunitx}
\usepackage{flafter}
\usepackage{placeins}
\title{Asgard/NOTT: Status of laboratory nulling performance}
\usepackage[left=0.76in,right=0.76in,top=1.00in,bottom=1.94in]{geometry}

\author[a]{T. Mattheussen (thomas.mattheussen@kuleuven.be)}
\author[a]{R. Laugier}
\author[a]{D. Defrère}
\author[b]{G. Garreau}
\author[a]{P. Chingaipe}
\author[a]{G. Raskin}
\author[a]{M. Salman}
\author[a]{K. Missiaen}
\author[a]{B. Vandenbussche}
\author[a]{W. Verstraeten}
\author[a]{J. Morren}
\author[a]{A. Kuriakose}
\author[c]{M. Ireland}
\author[d]{M.-A. Martinod}
\author[e]{A. Sanny}
\author[i]{S. Gross}
\author[l]{L. Labadie}
\author[a]{A. Bigioli}
\author[j]{A. Joó}
\author[c]{S. Madden}
\author[m]{A. Mazzoli}
\author[j]{G. Medgyesi}
\author[g,h]{S. Ertel}
\author[g]{T. A. Stuber}
\author[g]{J. P. Scott}
\author[k]{S. Kraus}
\author[d]{F. Martinache}
\author[f]{X. Haubois}
\author[f]{N. Schuhler}

\affil[a]{Institute of Astronomy, KU Leuven, Celestijnenlaan 200D, 3001 Heverlee, Belgium}
\affil[b]{ETH Zürich, Institute for Particle Physics \& Astrophysics, Wolfgang-Pauli-Str. 27, 8093 Zürich, Switzerland}
\affil[c]{Research School of Astronomy \& Astrophysics, Australian National University, Canberra, ACT 2611, Australia}
\affil[d]{Université Côte d’Azur, Observatoire de la Côte d’Azur, CNRS, Laboratoire Lagrange, France}
\affil[e]{European Southern Observatory, Garching, Germany}
\affil[f]{European Southern Observatory, Alonso de Cordova 3107 Vitacura, 19001 Santiago, Chile}
\affil[g]{Department of Astronomy and Steward Observatory, The University of Arizona, 933 North Cherry Avenue, Tucson, AZ 85721, USA}
\affil[h]{Large Binocular Telescope Observatory, The University of Arizona, 933 North Cherry Avenue, Tucson, AZ 85721, USA}
\affil[i]{MQ Photonics Research Centre, School of Mathematical and Physical Sciences, Macquarie University, NSW, 2109, Australia}
\affil[j]{Konkoly Observatory, 	HUN-REN Research Centre for Astronomy and Earth Sciences, Budapest, Hungary}
\affil[k]{Department of Physics and Astronomy, University of Exeter, Stocker Road, Exeter EX4 4QL, UK}
\affil[l]{I. Physikalisches Institut, Universität zu Köln, Zülpicher Str. 77, 50937 Köln, Germany}
\affil[m]{STAR Institute, University of Liège, 19C allée du Six Août, 4000 Liège, Belgium}

\begin{document} 
%\pagebreak
%\hspace{0pt}
%\vfill
%\begin{center}
%    Copyright [2026] Society of Photo‑Optical Instrumentation Engineers (SPIE). One print or electronic copy may be made for personal use only. Systematic reproduction and distribution, duplication of any material in this paper for a fee or for commercial purposes, or modification of the content of the paper are prohibited.
%\end{center}
%\vfill
%\hspace{0pt}
%\pagebreak

\maketitle
\begin{abstract}
Nulling interferometry enables the direct detection of faint companions and circumstellar structures at angular separations unresolvable by classical, diffraction-limited imagers, whilst dramatically improving the measurable contrast. The Asgard/NOTT nulling instrument aims to achieve a contrast performance of $10^{-5}$ in the L' wavelength band (\qty{3.5}-\qty{4.0}{\um}), enabling observation and characterization of young giant exoplanets near the snowline and hot exozodiacal dust. Previous studies have verified the nulling capabilities, of the chip in ambient conditions and of the test bench in cryogenic conditions. This work aims to add the first ambient performance assessment of the test bench with spectrally dispersed light. Necessary revisions are made to the data acquisition and calibration pipeline and fringe scans are carried out, modeled and fitted. The splitting ratios of the 4-telescope nulling beam combiner, a photonic Gallium Lanthanum Sulfide (GLS) chip, are moreover characterized on the bench, showing tentative agreement with previous chip characterization. The null performance has worsened, the achieved contrast of $\sim 10^{-1}$ being one order of magnitude higher than what earlier characterized performance showed. Multiple future changes to the test bed and to the approach taken promise an improved characterization of performance. In particular, the input beam intensities will be deliberately mismatched to account for the imbalanced splitting ratios of the directional couplers. With the installation of the final cryostat and camera, the developed tools will be leveraged to re-assess the performance in ambient and cryogenic conditions.
\end{abstract}


\section{INTRODUCTION}
\label{sec:intro}  
The observation and characterization of young giant exoplanets represents a crucial, yet observationally challenging step towards understanding planetary formation and evolution. Radial velocity surveys have uncovered a peak in the giant planet occurrence rate near the snowline \cite{Fernandes2019, Fulton2021}, where formation by core accretion is thought to be most efficient. The sensitivity and angular resolution required make imaging of planets in this region a challenging task for classical direct imagers \cite{Nielsen2019, Vigan2021}. Nulling interferometry, first proposed by Bracewell in 1978\cite{Bracewell1978}, achieves high-contrast measurements in the diffraction limit of classical direct imagers by establishing and maintaining a precise $\pi$ phase shift between collected beams before combination. As such, on-axis starlight is suppressed, enabling detection and characterization of off-axis light from faint companions and circumstellar structure. The development and commissioning of ground-based nulling interferometers will pave the way scientifically and technologically for future space-based nullers, meant to eventually characterize the atmospheres of rocky exoplanets around nearby stars, a rapidly growing \cite{Anglada2016, Gillon2017, Bonfils2018, Ribas2018, Damasso2020, Gonzalez2024, Nani2025} sample currently inaccessible for direct spectroscopic studies.

The NOTT instrument\cite{Defrere2022, Defrere2024} is part of the Asgard instrumental suite\cite{Martinod2023JATIS}, which saw first light in September 2025\cite{Ireland2026Asgard, Martinache2026_Heimdallr, Martinod2026Asgard, Kraus2026_BIFROST_status, Chhabra2026_BIFROST_LR}. Being the first long-baseline nuller for the Very Large Telescope Interferometer (VLTI) at the European Southern Observatory (ESO) Paranal site in Chile, its development, integration and operation will act as a pathfinder for future instruments like the Large Interferometer For Exoplanets (LIFE) mission\cite{Glauser2026LIFE}. NOTT will moreover be the first nuller to probe the L' (\qty{3.5}-\qty{4.0}{\um}) wavelength band, in which the planet-star contrasts are favourable ($\sim{10^{-5}}$) and the thermal background is fainter than for longer wavelengths. NOTT will leverage the Asgard/HEIMDALLR fringe tracker\cite{Martinache2026_Heimdallr}, the state-of-the-art adaptive optics at the VLTI\cite{NAOMI2019, GRAVITY+2026}, the maturing of integrated optics beam combiners in the mid-infrared, and novel post-processing data reduction techniques to meet this contrast requirement. In doing so, it has the potential to detect and characterize young giant exoplanets near the snowline \cite{Laugier2023} and hot exozodiacal dust\cite{Kral2017, Ertel2025, Scott2026exozodi}. 

\begin{figure}[t] 
    \centering
    \includegraphics[width=0.60\textwidth]{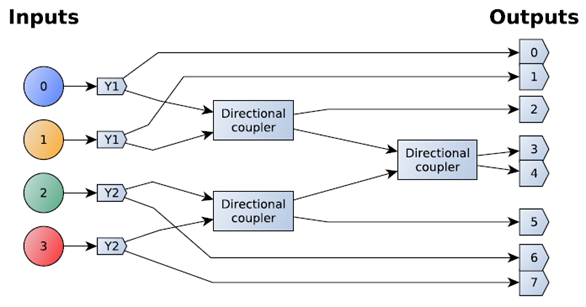}
    \caption[Schematic of the double Bracewell photonic beam combiner architecture used in NOTT.]{Schematic of the double Bracewell photonic beam combiner architecture used in NOTT, including the four inputs (four VLTI ATs/UTs or lab sources) of the chip, photometric taps Y1 and Y2, two combination stages (facilitated by directional couplers (DC)) and the outputs. The input beam phases are tuned such that the rightmost DC combines the dark (nulled) single-stage outputs of the previous two couplers to yield two complementary dark outputs (3, 4). The other outputs include the single-stage bright outputs (2, 5) and the non-coupled, isolated photometric outputs (0, 1, 6, 7). Figure adopted from Ref. \protect\citenum{Laugier2023}.}
    \label{beam_combiner}
\end{figure}

To verify the nulling capabilities of the instrument, a test bench was built at KU Leuven \cite{Garreau2024a, GarreauPhD}. The four beams are combined in a Gallium Lanthanum Sulfide (GLS) photonic chip\cite{Gretzinger19, SannyPhD}. As the written waveguides approach one another, the evanescent fields couple and the light interferes. This beam combination is performed in a Double Bracewell \cite{Angel1997, Mennesson2005} architecture, coupling the outputs of two first-stage directional couplers (DC) once more to achieve necessary robustness to instrumental errors in the differential output (see Fig. \ref{beam_combiner}). The 4-telescope nulling beam combiner is manufactured at Macquarie University using the ultrafast laser inscription (ULI) technique \cite{Sanny2026OL} and its nulling performance (contrast, stability, chromaticity) in ambient conditions is iteratively optimized\footnote{Several combination schemes, with different interaction lengths $l$, are written next to one another into the chip. The optimal scheme has $l =$ \qty{7.5}{\mm}.} for the (\qty{3.65}-\qty{3.85}{\um}) wavelength band using the characterization facility at the University of Cologne \cite{Sanny2026}. Using the NOTT test bench, the first nulling measurements with this chip were carried out in $2025$\cite{garreau2026nott} in broadband and in cryogenic conditions. The work presented here complements this cryogenic characterization by providing the first dispersed null measurements in ambient conditions, assessing the latest performance of the bench.
Section \ref{sec:methodology} provides an overview of the test bench, emphasizing the developments since the characterization in Ref. \citenum{garreau2026nott}. It also outlines the pipeline of data acquisition and calibration and the approach to acquiring and modeling fringes, by which the null performance is assessed. In Sect. \ref{sec:results}, we report on the obtained dispersed beam combiner splitting ratios and the obtained null performance. Section \ref{sec:conclusions} then assesses the next steps in terms of integration on the bench and data processing. 

\section{METHODOLOGY}
\label{sec:methodology}

\subsection{Test bench}
\label{sec:test_bench}
\begin{figure} [t]
\begin{center}
\begin{tabular}{c} 
\includegraphics[height=6cm]{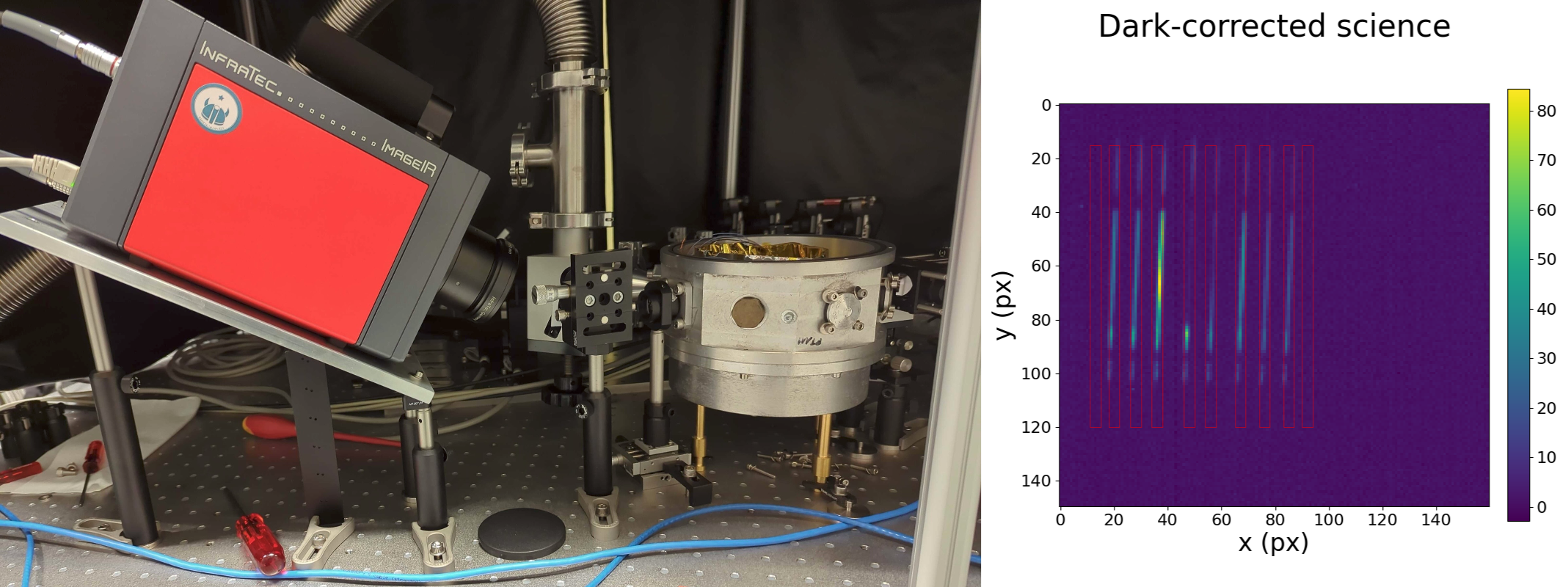}
\end{tabular}
\end{center}
\caption[example] 
{ \label{camera-tip} (left): The infrared dummy camera, the prism installed in a mount and the test cryostat. The camera is tipped to accommodate for the ray angle deviation after the prism, which happens in the vertical direction. (right): Dark- and background-subtracted readout of a double Bracewell null. Red boxes indicate the ten regions of interest (ROIs) defined on the detector frame. Eight of them are matched to the chip output channels, the other two are placed at opposite ends of the eight scientific ROIs and are used for background subtraction. The four photometric ROIs are denoted as P1, P2, P3 and P4, the four interferometric ROIs as I1, I2, I3 and I4. The background ROIs are denoted as B1 and B2. From left to right on the image, the ROIs are B2, P4, P3, I4, I3, I2, I1, P2, P1 and B1.
}
\end{figure}

A schematic overview of the test bench, as it was implemented in $2025$, is shown in Fig. \ref{testbench} \cite{garreau2026nott}. Developments since include:
\begin{figure} [t]
\begin{center}
\begin{tabular}{c} 
\includegraphics[height=12cm]{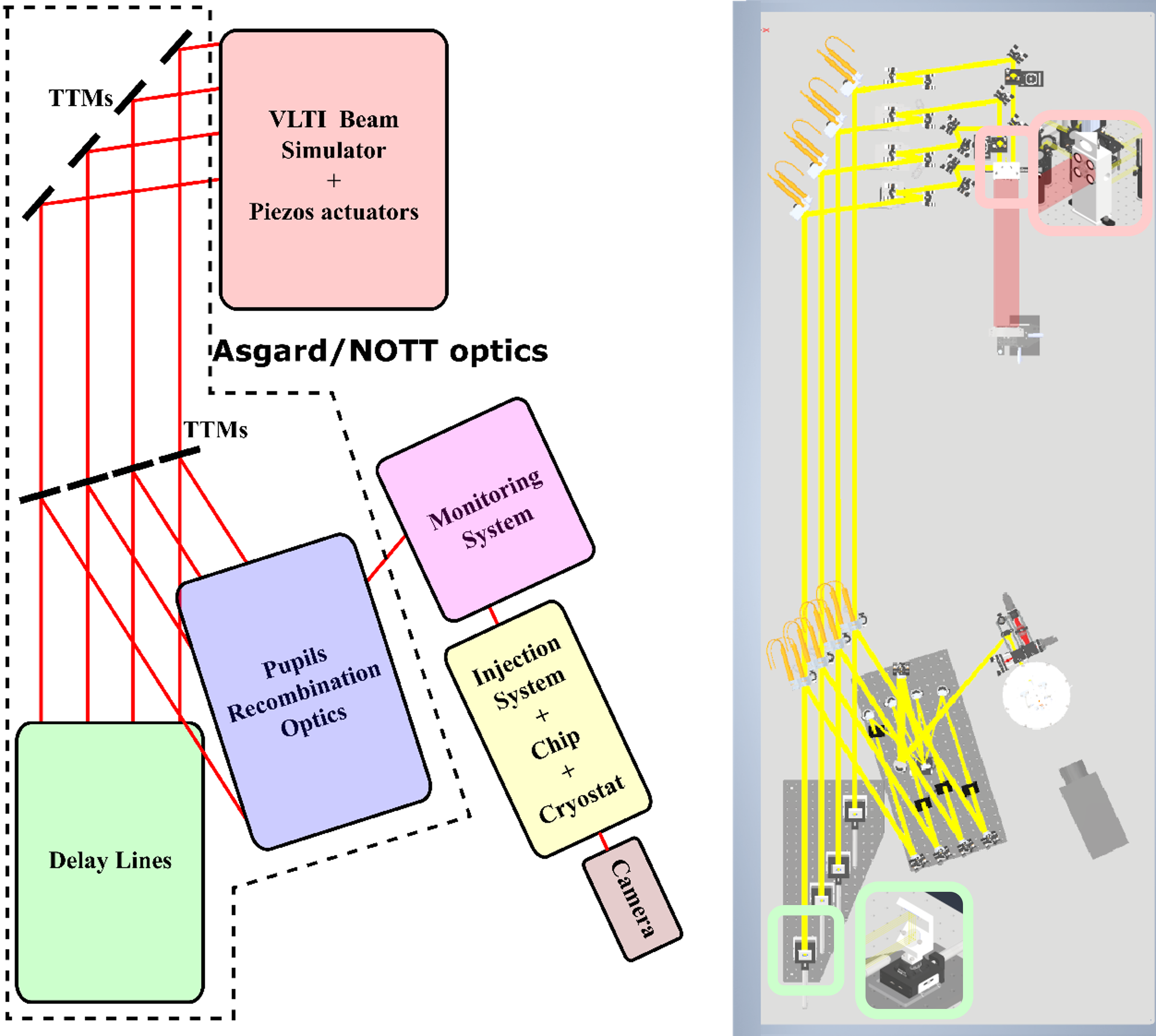}
\end{tabular}
\end{center}
\caption[example] 
{ \label{testbench} Schematic overview of the NOTT test bench, as implemented in $2025$ in the KU Leuven laboratory. Two light sources are available: a visible laser for alignment and a black body source with L' band filter for science. A simulator is manufactured to produce four collimated beams, resembling the beams that will be received from the VLTI telescopes. These beams pass through a sequence of optics to eventually reach the monitoring system, cryostat, the photonic chip beam combiner, the prism (not shown) and the infrared camera. The beams have a diameter of \qty{18}{\mm} at the beam simulator and inject into waveguides with a mode field diameter of about \qty{22}{\um} \protect\cite{Gretzinger19} at the photonic chip. Given this reduction of size, alignment of the beams in pupil and image plane is delicate. Two sets of tip-tilt mirrors (TTMs) allow for such alignment through coordinated motion. Visible cameras, integrated in the monitoring system, allow for follow-up of the beam positions. Delay line and piezo mirrors are integrated for respectively crude and precise cophasing of the beams. Refer to Ref. \protect\citenum{garreau2026nott} for a view of the piezo actuators and the test cryostat. Adapted with permission from Garreau et al., Proc. SPIE $14148$-$125$ $2026$.
}
\end{figure} 
\begin{itemize}

\item All four piezo actuators are motorized and operational, allowing for precise cophasing of the four beams.
\item The mounts of the tip-tilt mirrors were replaced as the previous Ultima U200-G2K gimbal models did not provide sufficient precision in the control of beam position to allow for the throughput criterion put forward in Ref. \citenum{Garreau2024JATIS}. New Siskyou IXF2.0a flexure mounts are installed, enhancing the precision. 
\item A prism is installed between the cryostat and the infrared camera. The prism and camera are aligned to eliminate vignetting and have uniform illumination onto the detector, see Fig. \ref{camera-tip}. 

\end{itemize}
Figure \ref{camera-tip} shows a double Bracewell null, obtained in the current state of the ambient test bench. It shows the entire wavelength bandpass of the lab source, spanning \qty{450}{\nm} to \qty{5500}{\nm}. The photometric outputs $0,1$ (P1, P2) and $6,7$ (P3, P4) reach mutually balanced levels of respectively $25$ and $40$ Digital Values (DV), after dark and background subtraction. Non-uniform illumination of the parabolic mirror in the beam simulator and remaining vignetting in the spectrograph are likely responsible for this mismatch. Since acquiring the data presented in this manuscript, the Longitudinal Dispersion Correctors\cite{Laugier2024LDC} (LDCs) have been installed, in between the two sets of TTMs. The cryogenic spectrograph and its Hawaii-2RG detector are being integrated in Leuven in July 2026. 

\subsection{Data acquisition and calibration}
\label{sec:data_acq_cal}
\begin{figure} [t]
\begin{center}
\begin{tabular}{c} 
\includegraphics[width=17cm]{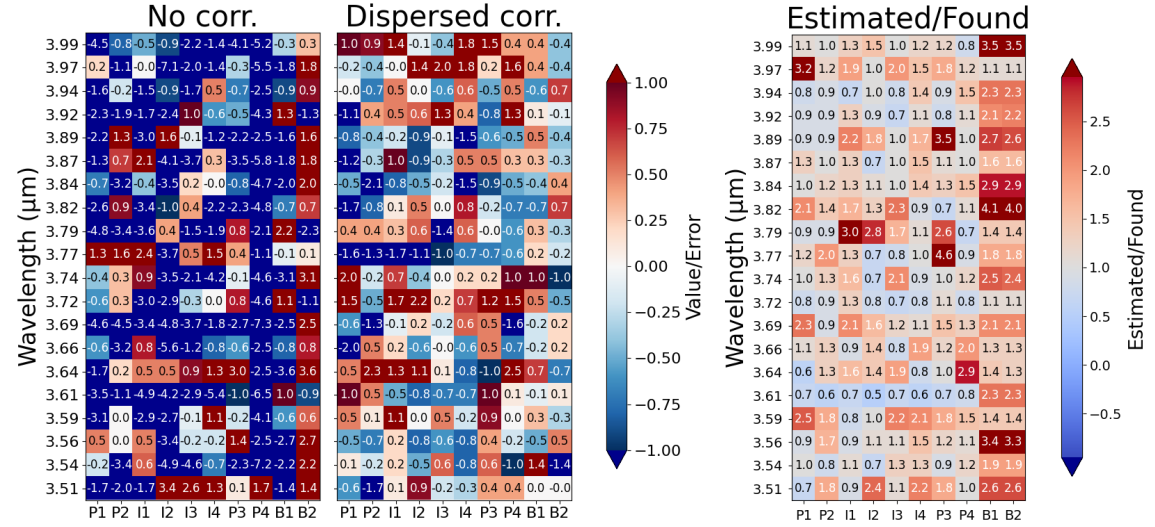}
\end{tabular}
\end{center}
\caption[example] 
{ \label{null_test} (left): Output of the null test, before (No corr.) and after (Dispersed corr.) correction of the spatial non-uniformity in the background. A raw background frame is given to the calibration pipeline as input. The ratio of the calibrated value and propagated error (the residual) is shown, at each wavelength and for all ROIs. After correction, the majority of residuals remain buried within the noise, showing a ratio within $[-1,1]$. (right): Output of the error test for a sequence of five calibrated master frames, spanning one second each. The ratio of the estimated and found variation is shown, at each wavelength and for all ROIs. The found error is underestimated for the background ROIs as the common modes of noise cancel upon subtraction of the mean background level. This is an artifact of the approach and does not affect the validity of the calibration when applied to the scientific ROIs.
}
\end{figure}

Given the addition of the prism, a new pipeline was built to acquire and calibrate windowed detector frames, identifying regions of interest (ROIs) post-acquisition and securing the wavelength structure probed by the pixel rows. Using a filter, the pixel rows that mark the boundaries of the L' wavelength band are determined. Wavelength calibration is performed by linear interpolation within these boundaries. The calibration pipeline subtracts a master dark frame and real-time background levels (ROI- and wavelength-specific) from each science frame in an acquired sequence. It then outputs calibrated master science frames, averaging frames within a given timespan, and propagated error maps. It can also collapse the pixel columns to return dispersed readouts for each ROI. This feature is used to obtain the data in Sec. \ref{sec:scan_model}. The master dark is taken before science acquisition by closing the shutters. The background levels are estimated per frame. The two outer ROIs (see Fig. \ref{camera-tip}) are averaged and collapsed along the pixel columns. The resulting, spectrally dispersed average background level is corrected (dispersed corr. in Fig. \ref{null_test}) for the column-wise spatial non-uniformity of the background emission to give the dispersed background levels specific to each ROI. These levels are then subtracted from each ROI in the science frame. The non-uniformity is characterized prior to any readout by use of a master background frame, co-adding a 120s-long sequence of frames with the light source turned off. The relative deviation of the dispersed background level in each ROI, compared to the dispersed average level of the two outer ROIs, is then deduced from this master frame. Two tests were carried out to assess the performance of this pipeline and the noise sources present in the test bench.
\begin{itemize}
\item Null test: upon passing a raw background frame as input, the calibration pipeline should output a zero map, as the input frame did not contain any scientific light. The calibrated intensity values should fall within the propagated errors. The ratio of value and error, here denoted as the residual, should lie within $[-1,1]$. Figure \ref{null_test} shows the output of such null test, before and after aforementioned dispersed correction for the spatial non-uniformity in the background. It shows the residuals for all wavelengths and for each ROI. After correction, the majority of residuals are within unity.

\item Error test: the light source is turned on and a sequence of calibrated master frames and propagated error maps is acquired. The found variation among the master frames is then compared to the mean estimated variation in the propagated error maps. Figure \ref{null_test} shows the output of such error test. If all sources of noise in the bench were perfectly white and stationary, the ratio of estimated and found error would be one. The observed ratios deviate from one in both directions. Although tentative, needing a longer sequence of exposures, this points to the presence of noise sources that act on different timescales.

\end{itemize}
\newpage
The detector noise is white and stationary, as shown by a power spectrum of dark frames. On the contrary, a power spectrum of the background frames showed an increase in power of two orders of magnitude for frequencies below $0.1$ Hz. On the timescales of the performed error test, this will add variance to the found error. The background level is moreover known to drift differentially between the ROIs and to be subject to the beam cophasing. The static non-uniformity correction will thus induce bias. We aim to implement a chopping approach on the bench, using e.g. a chopper wheel. This will allow for immediate subtraction of the real-time background level in the scientific ROIs. Were this not possible, we will revise the positioning of the background ROIs as to minimize the effect that spatial non-uniformity has on the calibration. 

\subsection{Fringe scanning, modeling and fitting}
\label{sec:scan_model}
Fringe scans are performed to assess the null performance on the test bench. Injection is optimized by tip-tilt motion and both pairs of beams ($1/2$ and $3/4$) are cophased to their coherent interference region by the delay lines. The piezo mirrors are then actuated to bring each pair to the white-light fringe and to scan around it. One such scan is obtained for each single Bracewell combiner pair. The procedure is then repeated for the two pairs mutually, obtaining a scan of the double Bracewell scheme. In this manuscript, we limit ourselves to analyzing the single Bracewell pair of beams $3/4$ (see Fig. \ref{beam_combiner}), combining in the first-stage directional coupler with optimal interaction length of $l =$ \qty{7.5}{\mm}. We do still show the double Bracewell scan in Sec. \ref{sec:results} as a proof-of-concept. The single Bracewell model for the fringes is built starting from a general interference model for a directional coupler (DC). This starting point is derived in Appendix \ref{sec:chip_imb}. Adaptations and additions to this model include:
\begin{itemize}
    \item An additional parameter $\eta$ is added. It represents all unmodeled factors that limit the maximum coherent visibility from being realized. It encompasses effects like polarization mismatches or chromatic effects but does not encompass photometric mismatch or coupling geometry.
    \item Parameters $b_{sum} = b_{1}+b_{2}$ and $b_{diff} = b_{1}-b_{2}$ are added. $b_1$ and $b_2$ represent the possible bias (in DV) induced by the background subtraction in the two respective outputs ($1$ and $2$) of the directional coupler. These parameters thus inform on the validity of the approach to background subtraction.
    \item All scans are performed with piezo actuators that exhibit linear behaviour
    \begin{equation}
        \begin{aligned}
            \Delta \phi (x, \lambda) = \frac{2\pi a}{\lambda}\left(x-x_{ref}(\lambda)\right) + \Delta \phi_0(\lambda)
        \end{aligned}
    \end{equation}
    where $x_{ref}(\lambda)$ is the piezo position of the central fringe, $\Delta \phi_0(\lambda)$ an additional phase offset and $a$ a coefficient left to probe any mismatch between the actual and the reported positions of the piezos. 
    
\end{itemize}
The sum and difference of the fringes in the dark and bright outputs are modeled as shown in Eq. \ref{sumdiffrev}. $P_1$ and $P_2$ are the photometric intensities of the two beams injecting into the DC. $\mathcal{E}_i$ is the fraction of power in waveguide $i$ transmitted to the bar state, not coupling to the other waveguide. $P^{*} = \frac{P_1+P_2}{2}$ is the average photometric intensity and $\delta P = \frac{P_1-P_2}{P_1+P_2}$ is the photometric mismatch. The photometric intensities and bar power fractions are retrieved from the bench. Section \ref{sec:split} outlines this retrieval and Sect. \ref{sec:results} shows the resulting data. The such-obtained Y-junction split ratios are used to obtain the true input photometric intensities from the ones registered at chip output.

\begin{equation}
\label{sumdiffrev}
\begin{aligned}
    I_{sum} &= 2P^{*}\left[1+\eta\sin(\Delta \phi)\sqrt{1-\delta P^{2}}\left(\sqrt{\mathcal{E}_2(1-\mathcal{E}_1)}-\sqrt{\mathcal{E}_1(1-\mathcal{E}_2)}\right)\right] + b_{sum} \\
    I_{diff} &= 2P^{*}\left[-\delta P + \frac{\mathcal{E}_1P_1-\mathcal{E}_2P_2}{P^{*}}-\eta\sin(\Delta \phi)\sqrt{1-\delta P^{2}}\left(\sqrt{\mathcal{E}_2(1-\mathcal{E}_1)}+\sqrt{\mathcal{E}_1(1-\mathcal{E}_2)}\right)\right] + b_{diff}
\end{aligned}
\end{equation}

$I_{diff}$ relates to the fringe visibility and shows how efficiently the power of the input beams can be steered between the interferometric channels as we scan the piezos. It encapsulates three sources of loss: photometric mismatch ($\delta P$), coupling geometry ($\mathcal{E}_1, \mathcal{E}_2$) and everything else ($\eta$). The coupling geometry relates to the splitting ratios. An imbalance in the splits limits the degree to which the input beams can interfere. $I_{sum}$ encapsulates the degree by which the system deviates from the ideal, lossless \cite{hanot2011null} case. A hierarchical Bayesian Markov Chain-Monte Carlo (MCMC) approach is used for fitting the model to the fringes. All spectral channels are fit simultaneously, maximizing a single posterior. Parameter $a$ is the only parameter that is shared across all spectral channels, that is “global", as the mechanical piezo behaviour should not depend on wavelength. All other parameters are “local", i.e. they are specific to a wavelength bin. A Gaussian loglikelihood function is used and the posterior space is sampled by an affine-invariant Monte Carlo sampler, by use of the python emcee package. In a first stage, the phase parameters $a$ and $x_{ref}$ are bootstrapped. In a second stage, all parameters are fit simultaneously. A burn-in of $2500$ steps is discarded and the sampled chain of $2000$ steps is thinned by a factor $10$ to avoid autocorrelation.

\subsection{Beam combiner splitting ratios}
\label{sec:split}
The Y-junction and DC splitting ratios are acquired from four master frames, each closing all-but-one shutter and allowing only one beam to inject. Additional ROIs are defined for background-subtraction, in between the ROIs that are shown in Fig. \ref{camera-tip}. The splitting ratios are then computed for each master frame by weighing the registered intensities at the chip output. Taking the Y-junction splitting ratios as en example, this involves comparing the intensity registered in the one lit photometric channel to the total intensity registered in the three lit interferometric channels.

\section{RESULTS}
\label{sec:results}

\subsection{Splitting ratios}
\label{sec:results:splits}

\begin{figure} [t]
\begin{center}
\begin{tabular}{c} 
\includegraphics[width=10cm]{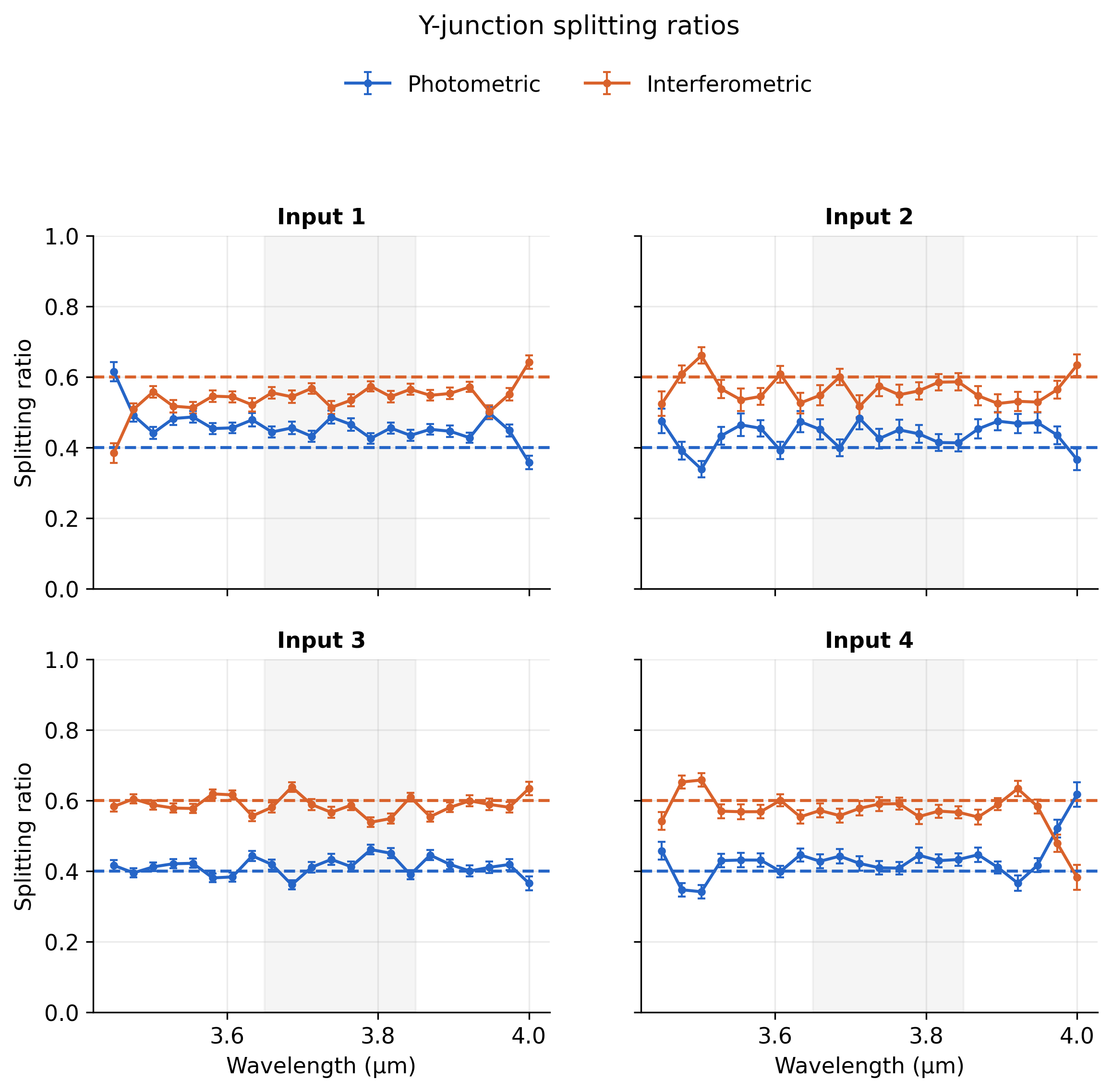}
\end{tabular}
\end{center}
\caption[example] 
{ \label{Y_junc_splits} The Y-junction photometric tap split ratios, for each input beam and across all wavelengths. Photometric denotes the fraction of input intensity that is passed to the photometric channel, interferometric the fraction passed to all interferometric channels. The shaded regions denote the wavelength band pass (\qty{3.65}-\qty{3.85}{\um}) used in Ref. \protect\citenum{SannyPhD, Sanny2026} to characterize and optimize the chip, the dashed lines indicate the splitting ratios found there.
}
\end{figure}

\begin{figure} [t]
\begin{center}
\begin{tabular}{c} 
\includegraphics[width=17cm]{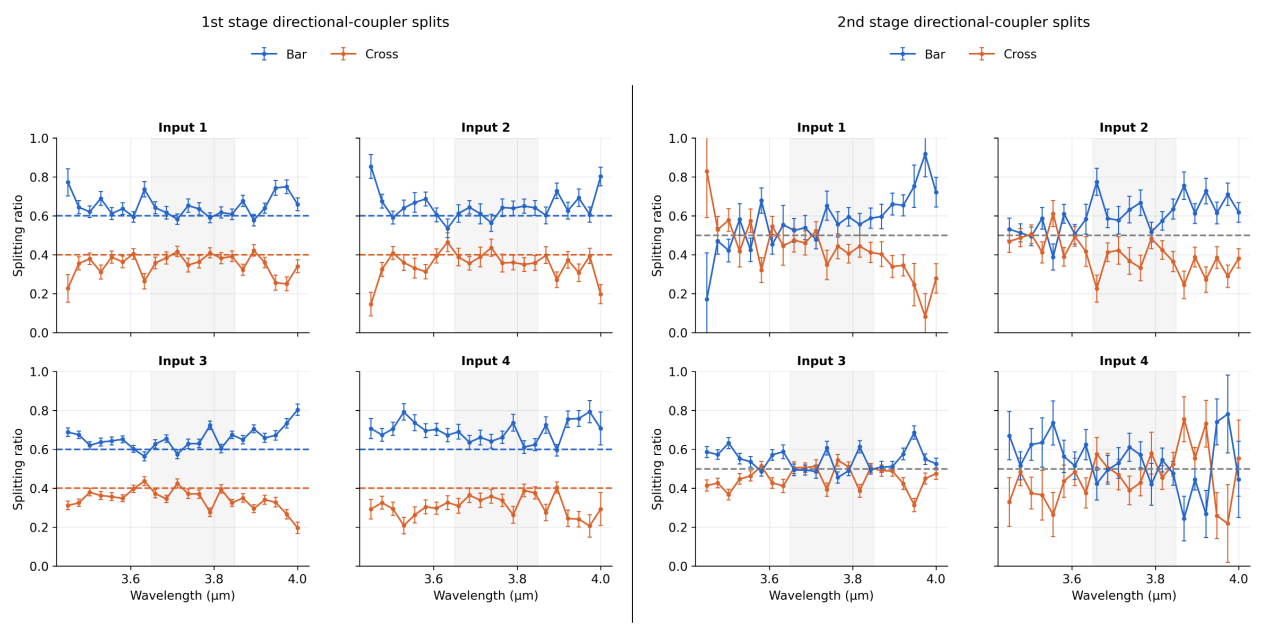}
\end{tabular}
\end{center}
\caption[example] 
{ \label{DC_1_splits} The first and second-stage DC split ratios, for each input beam and across all wavelengths. Bar denotes the fraction of input intensity that does not couple to the other waveguide. Cross denotes the fraction of input intensity that does. The shaded regions denote the wavelength band pass (\qty{3.65}-\qty{3.85}{\um}) used in Ref. \protect\citenum{SannyPhD, Sanny2026} to characterize and optimize the chip, the dashed lines indicate the splitting ratios found there.
}
\end{figure}

The splitting ratios are calculated by the approach outlined in Sect. \ref{sec:split}. Figs. \ref{Y_junc_splits} and \ref{DC_1_splits} respectively show the splitting ratios for the Y-junctions and the first- and second-stage DCs. The shaded regions denote the wavelength band pass (\qty{3.65}-\qty{3.85}{\um}) used in Ref. \citenum{SannyPhD, Sanny2026} to characterize and optimize the chip, finding achromatic splits for the Y-junctions and first-stage couplers ($60/40$) as well as for the second-stage coupler ($50/50$). The data presented here confirms that the splits are symmetric among each input pair ($1/2$ and $3/4$) within this wavelength band, as also found in Ref. \citenum{Sanny2026}. The data moreover confirms the achromaticity (within noise) in the L' band. This is the first time this result is confirmed on the test bench. The data however does not exactly align with the split ratios, most notably for DC3. This discrepancy worsens outside of the \qty{3.65}-\qty{3.85}{\um} band, as postulated in \citenum{GarreauPhD}. It must be noted that, before taking this data, it was not explicitly verified that the light injects into the optimal (interaction length $l =$ \qty{7.5}{\mm}\cite{Sanny2026}) nuller nor that the photonic chip and injection lens are aligned as to have uniform lens transmission onto the detector surface for all eight outputs. Although the system is stable, this will be verified before next measurements.
\subsection{Fringe scans}
\label{sec:results:fringe_scans}

\begin{figure} [t]
\begin{center}
\begin{tabular}{c} 
\includegraphics[width=15cm]{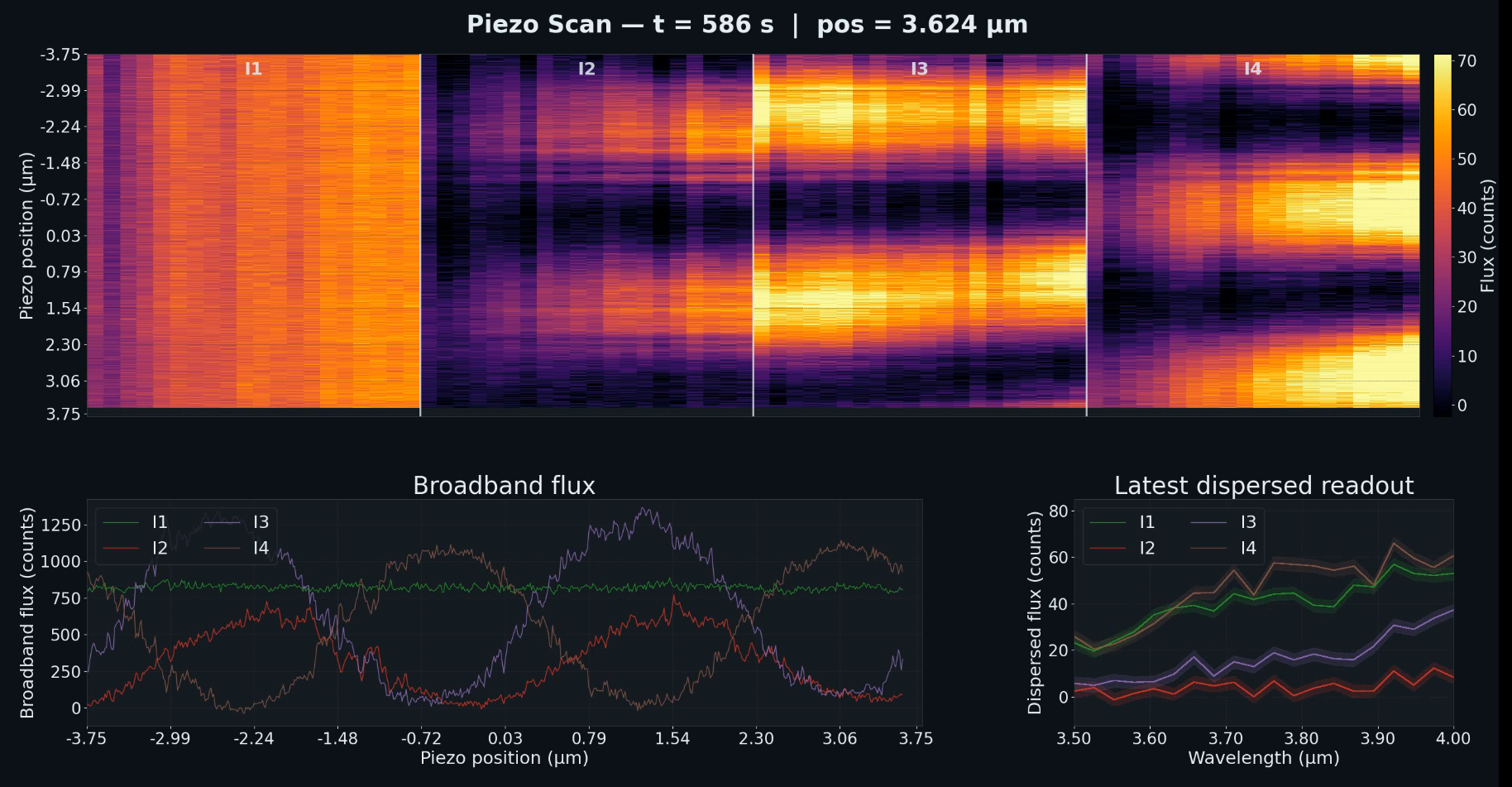}
\end{tabular}
\end{center}
\caption[example] 
{ \label{fig:double_bracewell_scan} Video: double\_bracewell\_scan. Scan of the double Bracewell null by pistoning the piezo mirror in beam channel $3$. The top panel shows the calibrated intensity in all four interferometric channels. The columns indicate wavelength, spanning the L' band. The bottom left panel shows the simultaneous broadband fluxes. The bottom right panel shows the latest dispersed readout. \url{http://dx.doi.org/doi.number.goes.here}}
\end{figure}

Figure \ref{fig:double_bracewell_scan} shows a fringe scan for the double Bracewell null configuration. Cophasing was performed manually and is therefore sub-optimal, as can be noted by the phase mismatch between the nulls of interferometric outputs $2$ and $3$. This will be improved by future automation scripts.

Appendix B. shows the posterior distributions for wavelength \qty{3.715}{\um}, for a scan of the single Bracewell combiner of inputs $3/4$. The fringe model was simplified, omitting parameter $\Delta \phi_0(\lambda)$ and fixing $a=1$ for the sake of convergence time. The background biases $b_{diff}$ and $b_{sum}$ show non-zero values, verifying that the approach to background subtraction is incomplete, as postulated in Sec. \ref{sec:data_acq_cal}. 

\begin{figure} [t]
\begin{center}
\begin{tabular}{c} 
\includegraphics[width=14.0cm]{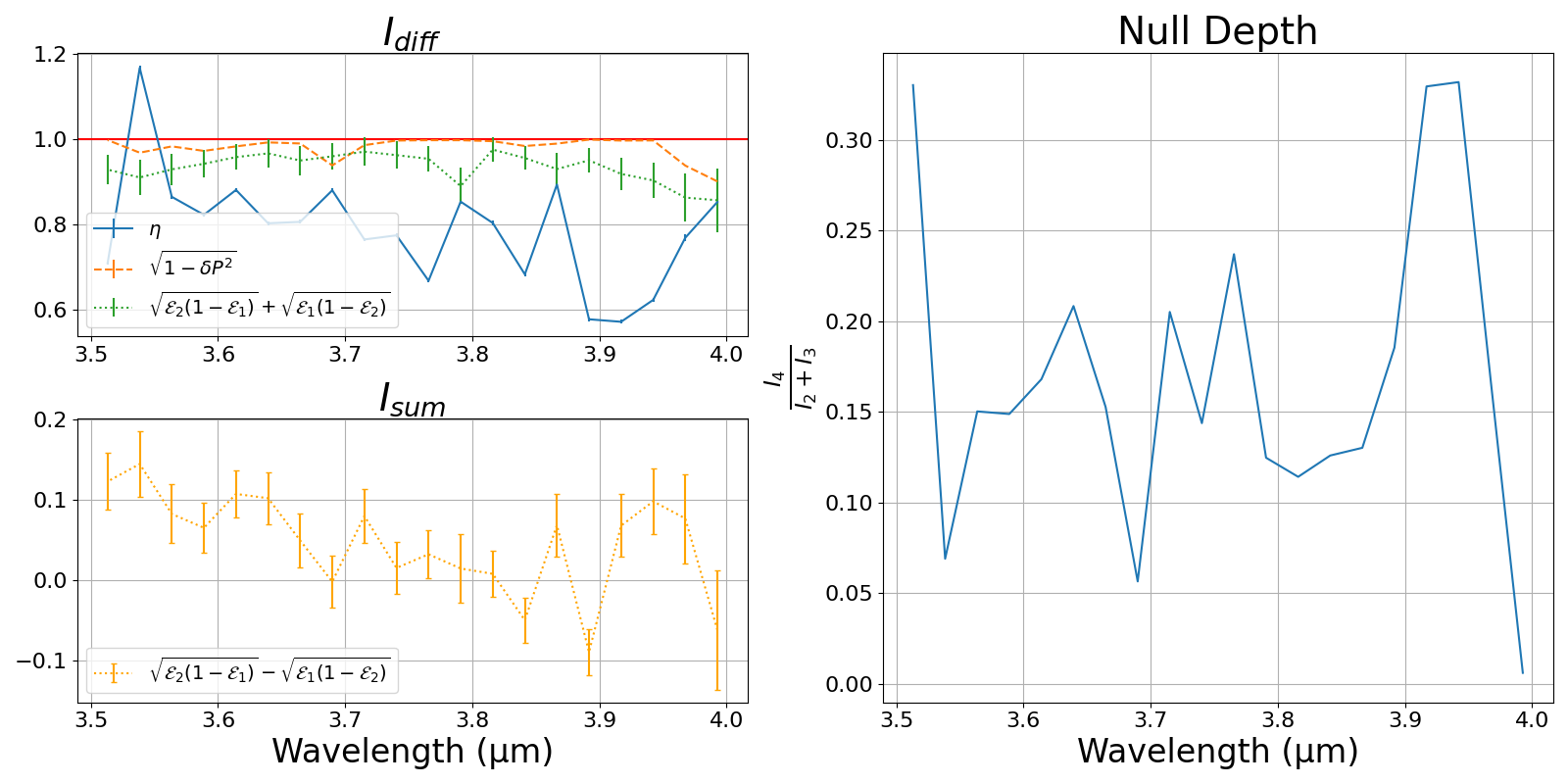}
\end{tabular}
\end{center}
\caption[example] 
{ \label{fig:vis_null} (top left): the visibility-limiting factors in the $I_{diff}$ signal. The photometric mismatch and coupling geometry factor are obtained from on-bench characterization, see Sect. \ref{sec:split}. The remaining $\eta$ factor is obtained from the MCMC analysis. As there are residual uncertainties in the determination of the splitting ratios this $\eta$ factor is allowed to exceed one. Values significantly above one are taken to flag model incompleteness rather than a physical result. (bottom left): the deviation from ideal coupling\protect\cite{hanot2011null} in the $I_{sum}$ signal. (right): the retrieved dispersed null depth.}
\end{figure}

The top-left panel in figure \ref{fig:vis_null} shows the visibility-limiting factors in $I_{diff}$, for each wavelength. The photometric mismatch and coupling geometry factor are obtained from on-bench characterization, see Sect. \ref{sec:split}. The remaining $\eta$ factor is obtained from the MCMC analysis. The bottom-left panel shows the coupling geometry factor in $I_{sum}$ and the right panel shows the dispersed null depths retrieved from the MCMC analysis by evaluating the fringe model with the median values of the fit parameters. The visibility is dominantly limited by unmodeled factors, the photometric mismatch and coupling geometry factors being close to one. The null depth does not match previously obtained performance \cite{Sanny2026, garreau2026nott}. It must however be noted that the intensities of the input beams were deliberately balanced in this work. Given the imbalanced split ratios of the first-stage DCs (see Fig. \ref{DC_1_splits}), this balance limits the achievable null depth. In future work, these intensities will be deliberately brought out of balance by tip-tilt motion to compensate for the imbalanced split ratios and minimize the null depth in the dark output. Lastly, we note that the modeled amplitude of the $I_{sum}$ signal tends to be underestimated compared to the data. Novel measurements for the splitting ratios will be carried out to assess whether this is an artifact of error in the characterized coupler geometry or an unmodeled physical artifact.

\section{CONCLUSIONS}
\label{sec:conclusions}
Verification of the nulling capabilities on the test bench marks an important step in the development of the Asgard/NOTT instrument. In this work, we set out to complement previous characterization in Refs. \citenum{Sanny2026} (chip: ambient, dispersed) and \citenum{garreau2026nott} (bench: cold, broadband) with the first dispersed characterization in ambient conditions on the test bench. Despite not achieving the same nulling performance, this work identifies the limiting factors and provides prospects for improvement to be implemented in next characterization. On the hardware side, the LDCs, the final Hawaii-2RG and the final cryostat have been installed. A novel assessment will be performed - before shipping to Paranal in August $2026$ - with the bench in its final state. Additional care will be given to the alignment of the chip-lens system, to guarantee injection into the optimized (interaction length $l =$ \qty{7.5}{\mm}) nuller and uniform transmission of the lens. We will moreover aim to either implement a chopper wheel to enable real-time background subtraction or to improve the approach taken here. The calibration pipeline will then be re-assessed in cryogenic conditions, tracing the origin of the 1/f drift in the background emission. Moreover, a mask has been manufactured to match the chip inputs, blocking stray light from passing through the chip, past the S-bend. In this final bench state, we will re-assess the split ratios and perform new fringe scans at ambient and cryogenic temperatures. The revised split ratios will inform the degree by which the intensities of the injecting beams need to be imbalanced. Deliberately imposing this imbalance, we aim to show dispersed null performance that matches previous characterizations of the system.
\FloatBarrier
\newpage

\acknowledgments % equivalent to \section*{ACKNOWLEDGMENTS}       
SCIFY has received funding from the European Research Council (ERC); Award no. CoG - 866070 under the European Union's Horizon 2020 research and innovation program. Ertel, S., Scott, J.P., and Stuber, T.A. acknowledge funding by the National Aeronautics and Space Administration (NASA) through grant 80NSSC23K1473. Martinod, M-A. and Martinache, F. wish to acknowledge funding from the project PHOTONICS financed by the ANR program PEPR Origins (ANR-22-EXOR-0005). Romain Laugier has received funding from the Research Foundation - Flanders (FWO) under the grant number 1234224N. Thomas Mattheussen has received funding from the Research Foundation - Flanders (FWO) under the grant number 1114226N.

% References
\bibliography{spie_proc_26} % bibliography data in report.bib

\begin{thebibliography}{10}

\bibitem{Fernandes2019}
{Fernandes}, R.~B., {Mulders}, G.~D., {Pascucci}, I., {Mordasini}, C., and {Emsenhuber}, A., ``{Hints for a Turnover at the Snow Line in the Giant Planet Occurrence Rate},'' {\em \apj}~{\bf 874},  81 (Mar. 2019).

\bibitem{Fulton2021}
Fulton, B.~J., Rosenthal, L.~J., Hirsch, L.~A., Isaacson, H., Howard, A.~W., Dedrick, C.~M., Sherstyuk, I.~A., Blunt, S.~C., Petigura, E.~A., Knutson, H.~A., Behmard, A., Chontos, A., Crepp, J.~R., Crossfield, I. J.~M., Dalba, P.~A., Fischer, D.~A., Henry, G.~W., Kane, S.~R., Kosiarek, M., Marcy, G.~W., Rubenzahl, R.~A., Weiss, L.~M., and Wright, J.~T., ``California legacy survey. ii. occurrence of giant planets beyond the ice line,'' {\em The Astrophysical Journal Supplement Series}~{\bf 255},  14 (jul 2021).

\bibitem{Nielsen2019}
Nielsen, E.~L., De~Rosa, R.~J., Macintosh, B., Wang, J.~J., Ruffio, J.-B., Chiang, E., Marley, M.~S., Saumon, D., Savransky, D., Mark~Ammons, S., Bailey, V.~P., Barman, T., Blain, C., Bulger, J., Burrows, A., Chilcote, J., Cotten, T., Czekala, I., Doyon, R., Duchêne, G., Esposito, T.~M., Fabrycky, D., Fitzgerald, M.~P., Follette, K.~B., Fortney, J.~J., Gerard, B.~L., Goodsell, S.~J., Graham, J.~R., Greenbaum, A.~Z., Hibon, P., Hinkley, S., Hirsch, L.~A., Hom, J., Hung, L.-W., Ilene~Dawson, R., Ingraham, P., Kalas, P., Konopacky, Q., Larkin, J.~E., Lee, E.~J., Lin, J.~W., Maire, J., Marchis, F., Marois, C., Metchev, S., Millar-Blanchaer, M.~A., Morzinski, K.~M., Oppenheimer, R., Palmer, D., Patience, J., Perrin, M., Poyneer, L., Pueyo, L., Rafikov, R.~R., Rajan, A., Rameau, J., Rantakyrö, F.~T., Ren, B., Schneider, A.~C., Sivaramakrishnan, A., Song, I., Soummer, R., Tallis, M., Thomas, S., Ward-Duong, K., and Wolff, S., ``The gemini planet imager exoplanet survey: Giant planet and brown dwarf demographics
  from 10 to 100 au,'' {\em The Astronomical Journal}~{\bf 158},  13 (jun 2019).

\bibitem{Vigan2021}
{Vigan}, A., {Fontanive}, C., {Meyer}, M., {Biller}, B., {Bonavita}, M., {Feldt}, M., {Desidera}, S., {Marleau}, G.-D., {Emsenhuber}, A., {Galicher}, R., {Rice}, K., {Forgan}, D., {Mordasini}, C., {Gratton}, R., {Le Coroller}, H., {Maire}, A.-L., {Cantalloube}, F., {Chauvin}, G., {Cheetham}, A., {Hagelberg}, J., {Lagrange}, A.-M., {Langlois}, M., {Bonnefoy}, M., {Beuzit}, J.-L., {Boccaletti}, A., {D'Orazi}, V., {Delorme}, P., {Dominik}, C., {Henning}, T., {Janson}, M., {Lagadec}, E., {Lazzoni}, C., {Ligi}, R., {Menard}, F., {Mesa}, D., {Messina}, S., {Moutou}, C., {M{\"u}ller}, A., {Perrot}, C., {Samland}, M., {Schmid}, H.~M., {Schmidt}, T., {Sissa}, E., {Turatto}, M., {Udry}, S., {Zurlo}, A., {Abe}, L., {Antichi}, J., {Asensio-Torres}, R., {Baruffolo}, A., {Baudoz}, P., {Baudrand}, J., {Bazzon}, A., {Blanchard}, P., {Bohn}, A.~J., {Brown Sevilla}, S., {Carbillet}, M., {Carle}, M., {Cascone}, E., {Charton}, J., {Claudi}, R., {Costille}, A., {De Caprio}, V., {Delboulb{\'e}}, A., {Dohlen}, K., {Engler}, N.,
  {Fantinel}, D., {Feautrier}, P., {Fusco}, T., {Gigan}, P., {Girard}, J.~H., {Giro}, E., {Gisler}, D., {Gluck}, L., {Gry}, C., {Hubin}, N., {Hugot}, E., {Jaquet}, M., {Kasper}, M., {Le Mignant}, D., {Llored}, M., {Madec}, F., {Magnard}, Y., {Martinez}, P., {Maurel}, D., {M{\"o}ller-Nilsson}, O., {Mouillet}, D., {Moulin}, T., {Orign{\'e}}, A., {Pavlov}, A., {Perret}, D., {Petit}, C., {Pragt}, J., {Puget}, P., {Rabou}, P., {Ramos}, J., {Rickman}, E.~L., {Rigal}, F., {Rochat}, S., {Roelfsema}, R., {Rousset}, G., {Roux}, A., {Salasnich}, B., {Sauvage}, J.-F., {Sevin}, A., {Soenke}, C., {Stadler}, E., {Suarez}, M., {Wahhaj}, Z., {Weber}, L., and {Wildi}, F., ``{The SPHERE infrared survey for exoplanets (SHINE). III. The demographics of young giant exoplanets below 300 au with SPHERE},'' {\em \aap}~{\bf 651},  A72 (July 2021).

\bibitem{Bracewell1978}
{Bracewell}, R.~N., ``{Detecting nonsolar planets by spinning infrared interferometer},'' {\em Nature}~{\bf 274},  780--781 (Aug. 1978).

\bibitem{Anglada2016}
{Anglada-Escud{\'e}}, G., {Amado}, P.~J., {Barnes}, J., {Berdi{\~n}as}, Z.~M., {Butler}, R.~P., {Coleman}, G. A.~L., {de La Cueva}, I., {Dreizler}, S., {Endl}, M., {Giesers}, B., {Jeffers}, S.~V., {Jenkins}, J.~S., {Jones}, H. R.~A., {Kiraga}, M., {K{\"u}rster}, M., {L{\'o}pez-Gonz{\'a}lez}, M.~J., {Marvin}, C.~J., {Morales}, N., {Morin}, J., {Nelson}, R.~P., {Ortiz}, J.~L., {Ofir}, A., {Paardekooper}, S.-J., {Reiners}, A., {Rodr{\'\i}guez}, E., {Rodr{\'\i}guez-L{\'o}pez}, C., {Sarmiento}, L.~F., {Strachan}, J.~P., {Tsapras}, Y., {Tuomi}, M., and {Zechmeister}, M., ``{A terrestrial planet candidate in a temperate orbit around Proxima Centauri},'' {\em Nature}~{\bf 536},  437--440 (Aug. 2016).

\bibitem{Gillon2017}
{Gillon}, M., {Triaud}, A. H.~M.~J., {Demory}, B.-O., {Jehin}, E., {Agol}, E., {Deck}, K.~M., {Lederer}, S.~M., {de Wit}, J., {Burdanov}, A., {Ingalls}, J.~G., {Bolmont}, E., {Leconte}, J., {Raymond}, S.~N., {Selsis}, F., {Turbet}, M., {Barkaoui}, K., {Burgasser}, A., {Burleigh}, M.~R., {Carey}, S.~J., {Chaushev}, A., {Copperwheat}, C.~M., {Delrez}, L., {Fernandes}, C.~S., {Holdsworth}, D.~L., {Kotze}, E.~J., {Van Grootel}, V., {Almleaky}, Y., {Benkhaldoun}, Z., {Magain}, P., and {Queloz}, D., ``{Seven temperate terrestrial planets around the nearby ultracool dwarf star TRAPPIST-1},'' {\em Nature}~{\bf 542},  456--460 (Feb. 2017).

\bibitem{Bonfils2018}
{Bonfils}, X., {Astudillo-Defru}, N., {D{\'\i}az}, R., {Almenara}, J.~M., {Forveille}, T., {Bouchy}, F., {Delfosse}, X., {Lovis}, C., {Mayor}, M., {Murgas}, F., {Pepe}, F., {Santos}, N.~C., {S{\'e}gransan}, D., {Udry}, S., and {W{\"u}nsche}, A., ``{A temperate exo-Earth around a quiet M dwarf at 3.4 parsec},'' {\em Astronomy and Astrophysics}~{\bf 613},  A25 (May 2018).

\bibitem{Ribas2018}
{Ribas}, I., {Tuomi}, M., {Reiners}, A., {Butler}, R.~P., {Morales}, J.~C., {Perger}, M., {Dreizler}, S., {Rodr{\'\i}guez-L{\'o}pez}, C., {Gonz{\'a}lez Hern{\'a}ndez}, J.~I., {Rosich}, A., {Feng}, F., {Trifonov}, T., {Vogt}, S.~S., {Caballero}, J.~A., {Hatzes}, A., {Herrero}, E., {Jeffers}, S.~V., {Lafarga}, M., {Murgas}, F., {Nelson}, R.~P., {Rodr{\'\i}guez}, E., {Strachan}, J.~B.~P., {Tal-Or}, L., {Teske}, J., {Toledo-Padr{\'o}n}, B., {Zechmeister}, M., {Quirrenbach}, A., {Amado}, P.~J., {Azzaro}, M., {B{\'e}jar}, V.~J.~S., {Barnes}, J.~R., {Berdi{\~n}as}, Z.~M., {Burt}, J., {Coleman}, G., {Cort{\'e}s-Contreras}, M., {Crane}, J., {Engle}, S.~G., {Guinan}, E.~F., {Haswell}, C.~A., {Henning}, T., {Holden}, B., {Jenkins}, J., {Jones}, H.~R.~A., {Kaminski}, A., {Kiraga}, M., {K{\"u}rster}, M., {Lee}, M.~H., {L{\'o}pez-Gonz{\'a}lez}, M.~J., {Montes}, D., {Morin}, J., {Ofir}, A., {Pall{\'e}}, E., {Rebolo}, R., {Reffert}, S., {Schweitzer}, A., {Seifert}, W., {Shectman}, S.~A., {Staab}, D., {Street}, R.~A.,
  {Su{\'a}rez Mascare{\~n}o}, A., {Tsapras}, Y., {Wang}, S.~X., and {Anglada-Escud{\'e}}, G., ``{A candidate super-Earth planet orbiting near the snow line of Barnard's star},'' {\em Nature}~{\bf 563},  365--368 (Nov. 2018).

\bibitem{Damasso2020}
{Damasso}, M., {Del Sordo}, F., {Anglada-Escud{\'e}}, G., {Giacobbe}, P., {Sozzetti}, A., {Morbidelli}, A., {Pojmanski}, G., {Barbato}, D., {Butler}, R.~P., {Jones}, H. R.~A., {Hambsch}, F.-J., {Jenkins}, J.~S., {L{\'o}pez-Gonz{\'a}lez}, M.~J., {Morales}, N., {Pe{\~n}a Rojas}, P.~A., {Rodr{\'\i}guez-L{\'o}pez}, C., {Rodr{\'\i}guez}, E., {Amado}, P.~J., {Anglada}, G., {Feng}, F., and {G{\'o}mez}, J.~F., ``{A low-mass planet candidate orbiting Proxima Centauri at a distance of 1.5 AU},'' {\em Science Advances}~{\bf 6},  eaax7467 (Jan. 2020).

\bibitem{Gonzalez2024}
{Gonz{\'a}lez Hern{\'a}ndez}, J.~I., {Su{\'a}rez Mascare{\~n}o}, A., {Silva}, A.~M., {Stefanov}, A.~K., {Faria}, J.~P., {Tabernero}, H.~M., {Sozzetti}, A., {Rebolo}, R., {Pepe}, F., {Santos}, N.~C., {Cristiani}, S., {Lovis}, C., {Dumusque}, X., {Figueira}, P., {Lillo-Box}, J., {Nari}, N., {Benatti}, S., {Hobson}, M.~J., {Castro-Gonz{\'a}lez}, A., {Allart}, R., {Passegger}, V.~M., {Zapatero Osorio}, M.~R., {Adibekyan}, V., {Alibert}, Y., {Allende Prieto}, C., {Bouchy}, F., {Damasso}, M., {D'Odorico}, V., {Di Marcantonio}, P., {Ehrenreich}, D., {Lo Curto}, G., {Santos}, R.~G., {Martins}, C.~J.~A.~P., {Mehner}, A., {Micela}, G., {Molaro}, P., {Nunes}, N., {Palle}, E., {Sousa}, S.~G., and {Udry}, S., ``{A sub-Earth-mass planet orbiting Barnard's star},'' {\em Astronomy and Astrophysics}~{\bf 690},  A79 (Oct. 2024).

\bibitem{Nani2025}
{Nari, N.}, {Dumusque, X.}, {Hara, N. C.}, {Suárez Mascareño, A.}, {Cretignier, M.}, {González Hernández, J. I.}, {Stefanov, A. K.}, {Passegger, V. M.}, {Rebolo, R.}, {Pepe, F.}, {Santos, N. C.}, {Cristiani, S.}, {Faria, J. P.}, {Figueira, P.}, {Sozzetti, A.}, {Zapatero Osorio, M. R.}, {Adibekyan, V.}, {Alibert, Y.}, {Allende Prieto, C.}, {Bouchy, F.}, {Benatti, S.}, {Castro-González, A.}, {D’Odorico, V.}, {Damasso, M.}, {Delisle, J. B.}, {Di Marcantonio, P.}, {Ehrenreich, D.}, {Génova-Santos, R.}, {Hobson, M. J.}, {Lavie, B.}, {Lillo-Box, J.}, {Lo Curto, G.}, {Lovis, C.}, {A. P. Martins, C. J.}, {Mehner, A.}, {Micela, G.}, {Molaro, P.}, {Mordasini, C.}, {Nunes, N.}, {Palle, E.}, {Quanz, S.P.}, {Ségransan, D.}, {Silva, A. M.}, {Sousa, S. G.}, {Udry, S.}, {Unger, N.}, and {Venturini, J.}, ``Revisiting the multi-planetary system of the nearby star hd 20794 - confirmation of a low-mass planet in the habitable zone of a nearby g-dwarf,'' {\em Astronomy and Astrophysics}~{\bf 693},  A297 (2025).

\bibitem{Defrere2022}
Defrère, D., Bigioli, A., Dandumont, C., Laugier, R., Garreau, G., Ireland, M., Berger, J., Courtney-Barrer, B., Loicq, J., and {More Authors}, ``L-band nulling interferometry at the vlti with asgard/hi-5: status and plans,'' in [{\em Optical and Infrared Interferometry and Imaging VIII}{\nolinebreak\hspace{0.1em}]},  M{\'e}rand, A., Sallum, S., and Sanchez-Bermudezv, J., eds., {\em Proceedings of SPIE} {\bf 12183}, SPIE, United States (2022).
\newblock SPIE Astronomical Telescopes + Instrumentation 2022 ; Conference date: 17-07-2022 Through 22-07-2022.

\bibitem{Defrere2024}
Defrère, D., Laugier, R., Martinod, M.-A., Garreau, G., Missiaen, K., Salman, M., Raskin, G., Dandumont, C., Loicq, J., and {More Authors}, ``L-band nulling interferometry at the vlti with asgard/nott: status and plans,'' in [{\em Optical and Infrared Interferometry and Imaging IX}{\nolinebreak\hspace{0.1em}]},  Kammerer, J., Sallum, S., and Sanchez-Bermudez, J., eds., {\em Proceedings of SPIE - The International Society for Optical Engineering} (2024).
\newblock SPIE Astronomical Telescopes + Instrumentation 2024 ; Conference date: 15-06-2024 Through 21-06-2024.

\bibitem{Martinod2023JATIS}
{Martinod}, M.-A., {Defr{\`e}re}, D., {Ireland}, M., {Kraus}, S., {Martinache}, F., {Tuthill}, P., {Bigioli}, A., {Bouzerand}, E., {Bryant}, J., {Chhabra}, S., {Courtney-Barrer}, B., {Crous}, F., {Cvetojevic}, N., {Dandumont}, C., {Ertel}, S., {Gardner}, T., {Garreau}, G., {Glauser}, A.~M., {Labadie}, L., {Lagadec}, T., {Laugier}, R., {Mazzoli}, A., {Mortimer}, D., {Norris}, B., {Raskin}, G., {Robertson}, G., {Sanny}, A., and {Taras}, A., ``{High-angular resolution and high contrast observations from Y to L band at the Very Large Telescope Interferometer with the Asgard Instrumental suite},'' {\em Journal of Astronomical Telescopes, Instruments, and Systems}~{\bf 9},  025007 (Apr. 2023).

\bibitem{Ireland2026Asgard}
Ireland, M.~J., Martinache, F., Kraus, S., Defr{\`e}re, D., Tuthill, P.~G., Ahrer, D.~J., Allouche, F., Bouzerand, E., Bryant, J.~J., Carter, J., Chhabra, S., Courtney-Barrer, B., Crous, F., Cvetojevic, N., Dahl, D.~S., Ertel, S., Gil, J.~P., Glauser, A.~M., Haubois, X., Labadie, L., Lagarde, S., Lancaster, D., Langford, C., Laugier, R., Ligi, R., McGinness, G., Martinod, M.-A., Mazzoli, A., Meilland, A., Missiaen, K.~D., Morel, S., Norris, B., Pallanca, L., Paul, J., Petrov, R.~G., Schuhler, N., Raskin, G., Robbe-Dubois, S., Robertson, G., Sanny, A., Snaith, O., and Taras, A.~K., ``Achieving the highest angular resolution and contrast with {Asgard} instrumental suite for {ESO}'s {VLTI}: commissioning results and projected performance,'' in [{\em SPIE Astronomical Telescopes + Instrumentation}{\nolinebreak\hspace{0.1em}]},   {\bf 14148}, SPIE (July 2026).

\bibitem{Martinache2026_Heimdallr}
Martinache, F., Ireland, M.~J., Taras, A.~K., Courtney-Barrer, B., Robbe-Dubois, S., Ligi, R., Cvetojevic, N., Petrov, R.~G., and Tuthill, P.~G., ``{Heimdallr}: integration and commissioning of the {VLTI/Asgard} fringe-tracker and {K}-band interferometric instrument,'' in [{\em Optical and Infrared Interferometry and Imaging X}{\nolinebreak\hspace{0.1em}]},  {\em Proc. SPIE} {\bf 14148},  14148--7 (July 2026).

\bibitem{Martinod2026Asgard}
Martinod, M.-A., Ireland, M.~J., Defr{\`e}re, D., Kraus, S., Martinache, F., Tuthill, P.~G., Bouzerand, E., Bryant, J.~J., Chhabra, S., Chingaipe, P., Courtney-Barrer, B., Crous, F., Cvetojevic, N., Dahl, D.~S., Dandumont, C., Ertel, S., Garreau, G., Glauser, A.~M., Haubois, X., Labadie, L., Lagarde, S., Lancaster, D., Laugier, R., Ligi, R., Long, N., Mazzoli, A., Mattheussen, T., Medgyesi, G., Missiaen, K.~D., Morel, S., Ahrer, D.~J., Norris, B., Paul, J., Petrov, R.~G., Raskin, G., Robbe-Dubois, S., Robertson, G., Salman, M., Sanny, A., Schuhler, N., Snaith, O., and Taras, A.~K., ``Pushing high angular resolution and high contrast observations on the {VLTI} from {Y} to {L} band with the {Asgard} instrumental suite: progress status and plans,'' in [{\em SPIE Astronomical Telescopes + Instrumentation}{\nolinebreak\hspace{0.1em}]},   {\bf 14148}, SPIE (July 2026).

\bibitem{Kraus2026_BIFROST_status}
Kraus, S., Lancaster, D., Chhabra, S., Paul, J., Snaith, O., Anugu, N., Monnier, J.~D., Ireland, M.~J., Taras, A.~K., Bianco, A., and Frangiamore, M., ``Opening short wavelengths for {VLTI}: {Asgard/BIFROST} project status and results from {AIV} and commissioning,'' in [{\em Optical and Infrared Interferometry and Imaging X}{\nolinebreak\hspace{0.1em}]},  {\em Proc. SPIE} {\bf 14148},  14148--9 (July 2026).

\bibitem{Chhabra2026_BIFROST_LR}
Chhabra, S., Kraus, S., Lancaster, D., Paul, J., Frangiamore, M., Bianco, A., Ireland, M.~J., and Snaith, O., ``Integration and commissioning of the {ASGARD/BIFROST} low-resolution ({R}~$\approx$~50--5000) spectrograph at the {VLTI},'' in [{\em Optical and Infrared Interferometry and Imaging X}{\nolinebreak\hspace{0.1em}]},  {\em Proc. SPIE} {\bf 14148},  14148--10 (July 2026).

\bibitem{Glauser2026LIFE}
Glauser, A.~M., Quanz, S.~P., Feinberg, L.~D., Agyemang, C., Alei, E., Besse, A., Birbacher, T., {van den Born}, J.~A., Bouzerand, E., Braam, M., Burr, Z., Dannert, F.~A., Dolkens, D., Eglin, D., Faist, J., Fortier, A., Garreau, G., Grange, R., Hansen, J., Hansen, J.~T., Huber, P.~A., Huisman, R., Ireland, M.~J., Kammerer, J., Kirchhoff, M., Kouwenhoven, K., Lai-Norling, J., Laugier, R., Meierhofer, R., Menti, F., Pitz, O., Satori, L.~F., Sigusch, I., Spalding, E., {van der Tak}, F., and {de Visser}, P., ``The {Large Interferometer For Exoplanets} ({LIFE}): establishing the foundations,'' in [{\em Optical and Infrared Interferometry and Imaging X}{\nolinebreak\hspace{0.1em}]},  {\em Proc.~SPIE} {\bf 14148}, SPIE, Copenhagen, Denmark (July 2026).
\newblock SPIE Astronomical Telescopes + Instrumentation 2026.

\bibitem{NAOMI2019}
{Woillez}, J., {Abad}, J.~A., {Abuter}, R., {Aller Carpentier}, E., {Alonso}, J., {Andolfato}, L., {Barriga}, P., {Berger}, J.-P., {Beuzit}, J.-L., {Bonnet}, H., {Bourdarot}, G., {Bourget}, P., {Brast}, R., {Caniguante}, L., {Cottalorda}, E., {Darr{\'e}}, P., {Delabre}, B., {Delboulb{\'e}}, A., {Delplancke-Str{\"o}bele}, F., {Dembet}, R., {Donaldson}, R., {Dorn}, R., {Dupeyron}, J., {Dupuy}, C., {Egner}, S., {Eisenhauer}, F., {Fischer}, G., {Frank}, C., {Fuenteseca}, E., {Gitton}, P., {Gont{\'e}}, F., {Guerlet}, T., {Guieu}, S., {Gutierrez}, P., {Haguenauer}, P., {Haimerl}, A., {Haubois}, X., {Heritier}, C., {Huber}, S., {Hubin}, N., {Jolley}, P., {Jocou}, L., {Kirchbauer}, J.-P., {Kolb}, J., {Kosmalski}, J., {Krempl}, P., {Le Bouquin}, J.-B., {Le Louarn}, M., {Lilley}, P., {Lopez}, B., {Magnard}, Y., {Mclay}, S., {Meilland}, A., {Meister}, A., {Merand}, A., {Moulin}, T., {Pasquini}, L., {Paufique}, J., {Percheron}, I., {Pettazzi}, L., {Pfuhl}, O., {Phan}, D., {Pirani}, W., {Quentin}, J., {Rakich}, A.,
  {Ridings}, R., {Riedel}, M., {Reyes}, J., {Rochat}, S., {Santos Tom{\'a}s}, G., {Schmid}, C., {Schuhler}, N., {Shchekaturov}, P., {Seidel}, M., {Soenke}, C., {Stadler}, E., {Stephan}, C., {Su{\'a}rez}, M., {Todorovic}, M., {Valdes}, G., {Verinaud}, C., {Zins}, G., and {Z{\'u}{\~n}iga-Fern{\'a}ndez}, S., ``{NAOMI: the adaptive optics system of the Auxiliary Telescopes of the VLTI},'' {\em \aap}~{\bf 629},  A41 (Sept. 2019).

\bibitem{GRAVITY+2026}
{Gravity+ Collaboration}, {Abuter}, R., {Allouche}, F., {Bailet}, C., {Benisty}, M., {Berdeu}, A., {Berger}, J.-P., {Berio}, P., {Bigioli}, A., {Blanchard}, C., {Boebion}, O., {Bonnet}, H., {Bourdarot}, G., {Bourget}, P., {Brandner}, W., {Brul{\'e}}, J., {Burgos}, P., {Carbillet}, M., {Correia}, C., {Courtney-Barrer}, B., {Curaba}, S., {Davies}, R., {Defr{\`e}re}, D., {Delboulb{\'e}}, A., {Delplancke}, F., {Dembet}, R., {Drescher}, A., {Dubost}, N., {Eckart}, A., {{\'E}douard}, C., {Eisenhauer}, F., {Esteras Otal}, L., {Fabricius}, M., {Feuchtgruber}, H., {F{\'e}dou}, P., {Finger}, G., {Schreiber}, N.~M.~F., {Frahm}, R., {Garcia}, E., {Garcia}, P., {Lopez}, R.~G., {Genzel}, R., {Gil}, J.~P., {Gillessen}, S., {Gomes}, T., {Gont{\'e}}, F., {Gopinath}, V., {Gouvret}, C., {Graf}, J., {Guajardo}, P., {Guieu}, S., {Hackenberg}, W., {Hartl}, M., {Haubois}, X., {Hau{\ss}mann}, F., {Henning}, T., {Hibon}, P., {H{\"o}nig}, S., {Horrobin}, M., {Houll{\'e}}, M., {Hubin}, N., {Taieb}, I.~I., {Jochum}, L., {Jocou}, L.,
  {Jost}, A., {Kammerer}, J., {Karl}, L., {Kaufer}, A., {Kern}, P., {Kervella}, P., {Kolb}, J., {Korhonen}, H., {Kreidberg}, L., {Krempl}, P., {Lacour}, S., {Lagarde}, S., {Lai}, O., {Lapeyr{\`e}re}, V., {Laugier}, R., {Leal}, V., {Le Bouquin}, J.-B., {Leftley}, J., {L{\'e}na}, P., {Lopez}, B., {Lutz}, D., {Magnard}, Y., {Mang}, F., {Marcotto}, A., {Maurel}, D., {M{\'e}rand}, A., {Millour}, F., {Montarges}, M., {More}, N., {Moruj{\~a}o}, N., {Moulin}, T., {Nowacki}, H., {Nowak}, M., {Oberti}, S., {Ott}, T., {Pallanca}, L., {Patru}, F., {Paumard}, T., {Perraut}, K., {Perrin}, G., {Petrucci}, P.~O., {Petrov}, R., {Pfuhl}, O., {Pourr{\'e}}, N., {Rabien}, S., {Rau}, C., {Riquelme}, M., {Robbe-Dubois}, S., {Rochat}, S., {Salman}, M., {S{\'a}nchez-Berm{\'u}dez}, J., {Schubert}, J., {Scigliuto}, J., {Shchekaturov}, P., {Schuhler}, N., {Shangguan}, J., {Shimizu}, T., {Scheithauer}, S., {Soenke}, C., {Soulez}, F., {Stadler}, E., {Stadler}, J., {Straubmeier}, C., {Sturm}, E., {Subroweit}, M., {Sykes}, C., {Tacconi},
  L.~J., {Tristram}, K.~R.~W., {Uysal}, S., {von Fellenberg}, S., {Widmann}, F., {Wieprecht}, E., {Wiezorrek}, E., {Woillez}, J., {Yazici}, S., and {Zins}, G., ``{First light for the GRAVITY+ Adaptive Optics: Extreme adaptive optics for the Very Large Telescope Interferometer},'' {\em \aap}~{\bf 707},  A115 (Mar. 2026).

\bibitem{Laugier2023}
{Laugier}, R., {Defr{\`e}re}, D., {Woillez}, J., {Courtney-Barrer}, B., {Dannert}, F.~A., {Matter}, A., {Dandumont}, C., {Gross}, S., {Absil}, O., {Bigioli}, A., {Garreau}, G., {Labadie}, L., {Loicq}, J., {Martinod}, M.-A., {Mazzoli}, A., {Raskin}, G., and {Sanny}, A., ``{Asgard/NOTT: L-band nulling interferometry at the VLTI. I. Simulating the expected high-contrast performance},'' {\em Astronomy and Astrophysics}~{\bf 671},  A110 (Mar. 2023).

\bibitem{Kral2017}
Kral, Q., Krivov, A.~V., Defrère, D., van Lieshout, R., Bonsor, A., Augereau, J.-C., Thébault, P., Ertel, S., Lebreton, J., and Absil, O., ``Exozodiacal clouds: hot and warm dust around main sequence stars,'' {\em Astronomical Review}~{\bf 13}(2),  69--111 (2017).

\bibitem{Ertel2025}
Ertel, S., Pearce, T.~D., Debes, J.~H., Faramaz, V.~C., Danchi, W.~C., Anche, R.~M., Defrère, D., Hasegawa, Y., Hom, J., Kirchschlager, F., Rebollido, I., Rousseau, H., Scott, J., Stapelfeldt, K., and Stuber, T.~A., ``Review and prospects of hot exozodiacal dust research for future exo-earth direct imaging missions,'' {\em Publications of the Astronomical Society of the Pacific}~{\bf 137},  031001 (mar 2025).

\bibitem{Scott2026exozodi}
{Scott}, J.~P., {Ertel}, S., {Stuber}, T.~A., {Defr{\`e}re}, D., and {Laugier}, R., ``{Performance analysis of the Asgard/NOTT nulling interferometer: optimizing observing modes for high-contrast detection},'' in [{\em Optical and Infrared Interferometry and Imaging X}{\nolinebreak\hspace{0.1em}]},  {\em Society of Photo-Optical Instrumentation Engineers (SPIE) Conference Series} {\bf 14148},  14148--70 (jul 2026).

\bibitem{Garreau2024a}
Garreau, G., Bigioli, A., Laugier, R., La~Torre, B., Martinod, M.-A., Missiaen, K., Morren, J., Raskin, G., Salman, M., Gross, S., Ireland, M.~J., Joó, A.~P., Labadie, L., Madden, S., Mazzoli, A., Medgyesi, G., Sanny, A., Taras, A., Vandenbussche, B., and Defrère, D., ``Asgard/nott: first lab assembly and experimental results,'' in [{\em Optical and Infrared Interferometry and Imaging IX}{\nolinebreak\hspace{0.1em}]},  Sallum, S., Sanchez-Bermudez, J., and Kammerer, J., eds.,  25, SPIE (Aug. 2024).

\bibitem{GarreauPhD}
{Garreau}, G., {\em Nulling interferometry: from exozodiacal dust with the LBTI to exoplanets with Asgard/NOTT at the VLTI}, PhD thesis, KU Leuven (2025).

\bibitem{Gretzinger19}
Gretzinger, T., Gross, S., Arriola, A., and Withford, M.~J., ``Towards a photonic mid-infrared nulling interferometer in chalcogenide glass,'' {\em Opt. Express}~{\bf 27},  8626--8638 (Mar 2019).

\bibitem{SannyPhD}
{Sanny}, A., {\em Towards the Spectroscopy of Giant Exoplanets with Nulling Interferometry: Fabrication \& Characterisation of the Mid-Infrared 4-Telescope Photonic Beam Combiner of NOTT/VLTI}, PhD thesis, Universit{\"a}t zu K{\"o}ln (2024).

\bibitem{Angel1997}
Angel, J. R.~P. and Woolf, N.~J., ``An imaging nulling interferometer to study extrasolar planets,'' {\em The Astrophysical Journal}~{\bf 475},  373 (jan 1997).

\bibitem{Mennesson2005}
{Mennesson}, A., {Borde}, P., {Boudet}, T., {Colavita}, M.~M., {Crawford}, S.~L., {Creech-Eakman}, M.~J., {Koresko}, C.~D., {Lipscy}, S.~J., {Milland-Gabet}, R., {Serabyn}, E., {Shao}, M., {Swanson}, P.~N., {van Belle}, G.~T., and {Vasisht}, G., ``{The Dusty AGB Star RS CrB: First Mid-Infrared Interferometric Observations with the Keck Telescopes},'' {\em \text{The Astrophysical Journal Letters}}~{\bf 634},  L169--L172 (Nov. 2005).

\bibitem{Sanny2026OL}
Sanny, A., Gretzinger, T., Gross, S., Labadie, L., and Withford, M., ``Mid-infrared nulling interferometry beam combiners using asymmetric directional couplers,'' {\em Optics Letters}~{\bf 51}(10),  2832--2835 (2026).

\bibitem{Sanny2026}
{Sanny}, A., {Labadie}, L., {Gross}, S., {Barjot}, K., {Laugier}, R., {Garreau}, G., {Martinod}, M.-A., {Defr{\`e}re}, D., and {Withford}, M.~J., ``{Asgard/NOTT: L-band nulling interferometry at the VLTI: III. The mid-infrared integrated optics beam combiner for NOTT},'' {\em \aap}~{\bf 705},  A37 (Jan. 2026).

\bibitem{garreau2026nott}
Garreau, G., Defr{\`e}re, D., Laugier, R., Chingaipe, P.~M., Martinod, M.-A., Mattheussen, T., Missiaen, K.~D., Morren, J., Raskin, G., Salman, M., Verstraeten, W., Bigioli, A., Ertel, S., and Gross, S., ``{Asgard/NOTT: cryogenic characterization of the mid-infrared photonic chip},'' in [{\em Optical and Infrared Interferometry and Imaging X}{\nolinebreak\hspace{0.1em}]},   {\bf 14148}, SPIE (July 2026).

\bibitem{Garreau2024JATIS}
{Garreau}, G., {Bigioli}, A., {Laugier}, R., {Raskin}, G., {Morren}, J., {Berger}, J.-P., {Dandumont}, C., {Goldsmith}, H.-D.~K., {Gross}, S., {Ireland}, M., {Labadie}, L., {Loicq}, J., {Madden}, S., {Martin}, G., {Martinod}, M.-A., {Mazzoli}, A., {Sanny}, A., {Shao}, H., {Yan}, K., and {Defr{\`e}re}, D., ``{Asgard/NOTT: L-band nulling interferometry at the VLTI. II. Warm optical design and injection system},'' {\em Journal of Astronomical Telescopes, Instruments, and Systems}~{\bf 10},  015002 (Jan. 2024).

\bibitem{Laugier2024LDC}
{Laugier}, R., {Defr{\`e}re}, D., {Ireland}, M., {Garreau}, G., {Absil}, O., {Matter}, A., {Petrov}, R., {Berio}, P., {Tuthill}, P., {Labadie}, L., {Martinod}, M.-A., and {Labadie}, L., ``{Asgard/NOTT: water vapor and CO$_{2}$ atmospheric dispersion compensation system},'' in [{\em Optical and Infrared Interferometry and Imaging IX}{\nolinebreak\hspace{0.1em}]},  {Kammerer}, J., {Sallum}, S., and {Sanchez-Bermudez}, J., eds., {\em Society of Photo-Optical Instrumentation Engineers (SPIE) Conference Series} {\bf 13095},  130952C (Aug. 2024).

\bibitem{hanot2011null}
Hanot, C., Mennesson, B., Martin, O., Serabyn, E., Denis, L., Irons, C., and Vasisht, G., ``Null self-calibration of nulling interferometers: theory and experiment,'' {\em Astronomy \& Astrophysics}~{\bf 531},  A34 (2011).

\end{thebibliography}
\bibliographystyle{spiebib} % makes bibtex use spiebib.bst
\FloatBarrier
\newpage
\appendix    %>>>> this command starts appendixes

\section{Directional coupler split imbalance}
\label{sec:chip_imb}
Waves of unequal amplitude can never perfectly cancel one another, hence an imbalance of the bar / cross power ratios in a DC will limit the achievable visibility in the outputs if not compensated by a mismatch of the input intensities.

Consider electric field wavefronts $E_1(\lambda)$ and $E_2(\lambda)$ entering the two input waveguides of a DC. The light can couple into the other waveguide, can “cross", picking up a $\frac{\pi}{2}$ phase shift in the process. We say the portion that does not couple into the other waveguide follows the “bar" path. Say waveguides $(1,2)$ have respective bar power fractions $(\mathcal{E}_{1},\mathcal{E}_{2})$ and cross power fractions $(1-\mathcal{E}_{1},1-\mathcal{E}_{2})$. Say moreover that $E_1$ has reference phase $\phi_1=0$ and $E_2$ has a relative phase offset $\Delta \phi = \phi_2 - \phi_1 = \phi_2$ with respect to $E_1$. The electric fields at the output of the DC can then be written as

\begin{equation}
\label{dc_outputs}
\begin{aligned}
E_{out,1} &= \sqrt{\mathcal{E}_1}E_1+i\sqrt{1-\mathcal{E}_2}E_2e^{i\Delta \phi} \\
E_{out,2} &= \sqrt{\mathcal{E}_2}E_2e^{i\Delta \phi}+i\sqrt{1-\mathcal{E}_1}E_1.
\end{aligned}
\end{equation}
The output intensities in waveguides $1$ and $2$ are then calculated as

\begin{equation*}
\label{I_out1}
\begin{aligned}
I_{out,1} &= \frac{1}{2}E_{out,1}^{*}E_{out,1} \\
&= \frac{1}{2}\left(\sqrt{\mathcal{E}_1}E_1+i\sqrt{1-\mathcal{E}_2}E_2e^{i\Delta \phi}\right)\left(\sqrt{\mathcal{E}_1}E_1-i\sqrt{1-\mathcal{E}_2}E_2e^{-i\Delta \phi}\right) \\
&= \frac{1}{2}\left[\mathcal{E}_1E_1^{2}+(1-\mathcal{E}_2)E_2^{2}-2\sqrt{\mathcal{E}_1}\sqrt{1-\mathcal{E}_2}E_1E_2\sin(\Delta \phi)\right] \\
&= \mathcal{E}_1P_1+(1-\mathcal{E}_2)P_2-2\sqrt{\mathcal{E}_1}\sqrt{1-\mathcal{E}_2}\sqrt{P_1P_2}\sin(\Delta \phi) \\
I_{out,2} &= \mathcal{E}_2P_2 + (1-\mathcal{E}_1)P_1+2\sqrt{\mathcal{E}_2}\sqrt{1-\mathcal{E}_1}\sqrt{P_1P_2}\sin(\Delta \phi)
\end{aligned}
\end{equation*}
with input photometric light intensities $P_1=\frac{1}{2}E_{in,1}^{*}E_{in,1}$ and $P_2=\frac{1}{2}E_{in,2}^{*}E_{in,2}$. Note how the case $\mathcal{E}_1 = \mathcal{E}_2=\frac{1}{2}$ recovers the ideal, lossless and perfectly symmetric coupler, as described in Ref. \citenum{hanot2011null}. 

Taking the sum and difference of the two output intensities we obtain:

\begin{equation}
\label{sumdiff}
\begin{aligned}
    I_{sum} &= P_1+P_2+2\sin(\Delta \phi)\sqrt{P_1P_2}\left[\sqrt{\mathcal{E}_2(1-\mathcal{E}_1)}-\sqrt{\mathcal{E}_1(1-\mathcal{E}_2)}\right] \\
    I_{diff} &= (2\mathcal{E}_1-1)P_1+(1-2\mathcal{E}_2)P_2-2\sin(\Delta \phi)\sqrt{P_1P_2}\left[\sqrt{\mathcal{E}_2(1-\mathcal{E}_1)}+\sqrt{\mathcal{E}_1(1-\mathcal{E}_2)}\right].
\end{aligned}
\end{equation}
Introducing the average photometric intensity $P^{*} = \frac {P_1+P_2}{2}$ and the photometric mismatch $\delta P = \frac{P_1-P_2}{P_1+P_2}$ gives
\begin{equation}
\label{sumdiff2}
\begin{aligned}
    I_{sum} &= 2P^{*}\left[1+\sin(\Delta \phi)\sqrt{1-\delta P^{2}}\left(\sqrt{\mathcal{E}_2(1-\mathcal{E}_1)}-\sqrt{\mathcal{E}_1(1-\mathcal{E}_2)}\right)\right] \\
    I_{diff} &= 2P^{*}\left[-\delta P + \frac{\mathcal{E}_1P_1-\mathcal{E}_2P_2}{P^{*}}-\sin(\Delta \phi)\sqrt{1-\delta P^{2}}\left(\sqrt{\mathcal{E}_2(1-\mathcal{E}_1)}+\sqrt{\mathcal{E}_1(1-\mathcal{E}_2)}\right)\right]
\end{aligned}
\end{equation}

\section{Single Bracewell posterior distributions}
\begin{figure} [H]
\begin{center}
\begin{tabular}{c} 
\includegraphics[width=17cm]{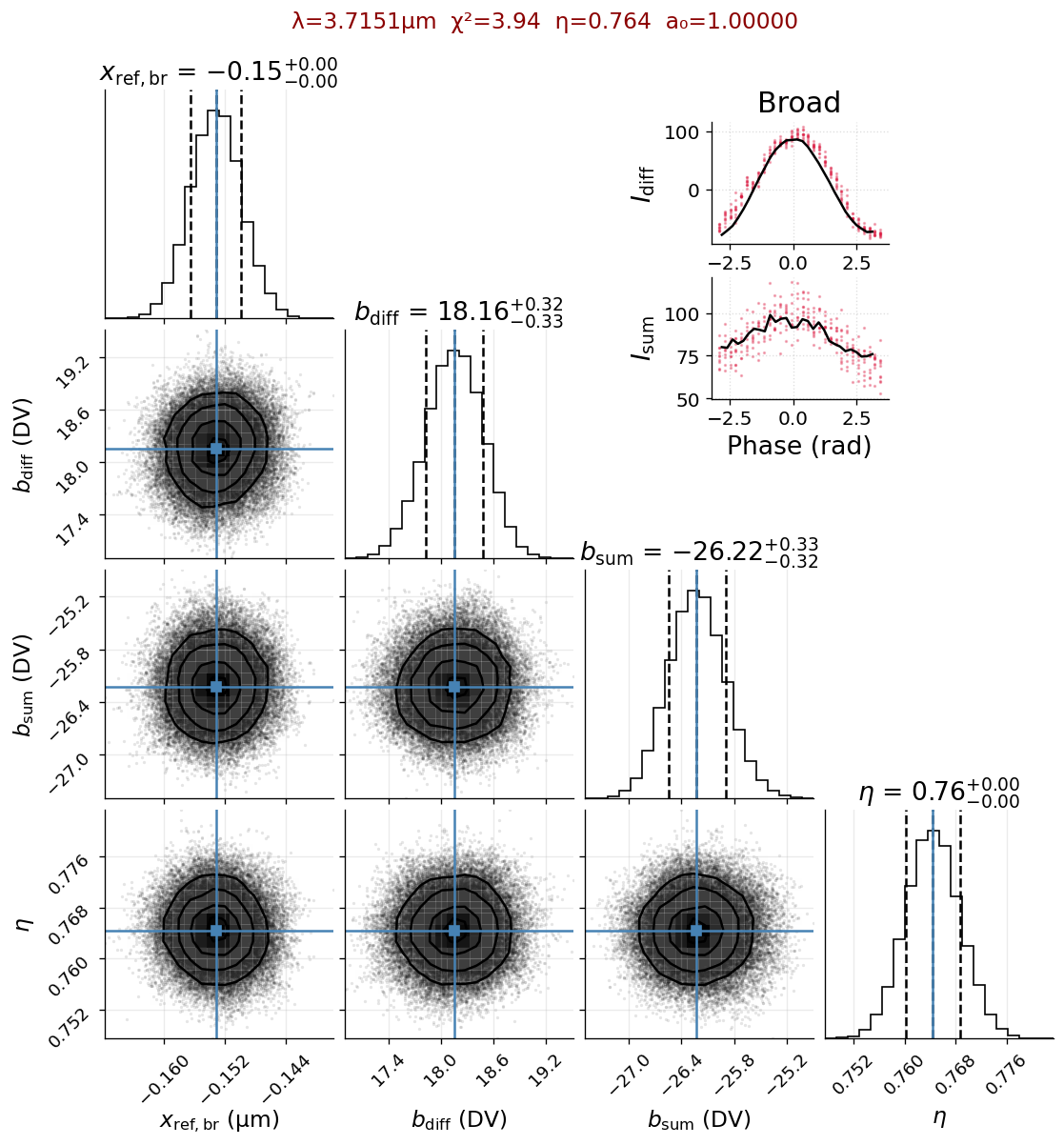}
\end{tabular}
\end{center}
\caption[example] 
{ \label{mcmc} (Main panel): posterior distributions of the fit parameters $x_{ref}$, $b_{diff}$, $b_{sum}$ and $\eta$ for wavelength \qty{3.715}{\um} of the single Bracewell scan for inputs $3/4$. The parameter $\Delta \phi(\lambda)$ is omitted and the coefficient $a$ is fixed to $1$ to simplify the model and aid convergence. The median is reported as fit value and the $16\%$ and $84\%$ quantiles are reported as errors. (Corner panel): $I_{diff}$ and $I_{sum}$ data alongside an evaluation of the fitted model. 
}
\end{figure} 

\end{document}